# Stochastic Power System Simulation Using the Adomian Decomposition Method

Nan Duan, *Student Member, IEEE*, and Kai Sun, *Senior Member, IEEE*

***Abstract*-- Considering increasing distributed energy resources and responsive loads in smart grid, this paper proposes a stochastic simulation approach for stability analysis of a power system having stochastic loads. The proposed approach solves a stochastic, nonlinear differential equation model of the system in an analytical way by the Adomian decomposition method and generates semi-analytical solutions that express both deterministic and stochastic state variables explicitly as symbolic variables so as to embed stochastic processes directly into the solutions for efficient stability analysis with uncertainties. The proposed approach is tested on the New England 10-machine 39-bus system with different penetration levels of stochastic loads. The approach is also benchmarked with a traditional stochastic simulation approach based on the Euler-Maruyama method. The results show that the new approach has better time performance and a comparable accuracy.**

## I. Introduction

UNCERTAINTIES exist in operations of power grids [1] Many factors such as random load consumptions and unanticipated relay protection actions contribute to the randomness of grid operations. It can be foreseen that a future power grid will have more uncertainties and stochastic behaviors in system operations due to the increasing penetrations of responsive loads and intermittent renewable generations. Thus, dynamic security assessment (DSA) of power systems should be conducted in both deterministic and stochastic manners. However, most of today's power system simulation software tools are still based on solvers of deterministic nonlinear, differential-algebraic equations (DAEs) that do not involve stochastic variables to model uncertainties in system operating conditions.

In literature, there are three major approaches for the modeling of a dynamic system having stochastic effects: the master equation, the Fokker-Planck equation [2][3] and Gillespie method [1][5]. The master equation and the Fokker-Plank equation are widely applied in the field of computational biology, which both focus on the evolution of probability distribution; the Gillespie method focuses on individual stochastic trajectories. The first two approaches provide a more comprehensive understanding of stochastic effects with a dynamic system but require solving high dimensional partial differential equations, so they are computationally difficult to be applied to simulations of realistic power systems [6]. There have been works using the Gillespie method for power system simulation [7]-[10].

In recent years, some researchers have contributed to power system simulation in a less-deterministic manner. Reference [7] proposed a systematic method to simulate the system behaviors under the influence of stochastic perturbations on loads, bus voltages and rotor speeds. This approach introduces stochastic differential equations (SDEs) to represent stochastic perturbations and solves the equations by Ito calculus, and then a mean trajectory with the envelope on trajectory variations is yielded by repeating simulations for many times. Papers [8]-[10] utilize a similar approach to study power system stability under random effects. To analyze long-term dynamics of a power system with wind generation, a new SDE model is developed in [11], which also applies the singular perturbation theory to investigate the slow dynamics of the system with stochastic wind generation. However, the time performance of such an approach based on Euler-Maruyama method can hardly meet the requirements for online power system simulation. Especially, when the penetration of distributed energy resources (DERs) reaches a high level, the distribution network behaves in a more stochastic manner as seen from the transmission network, and hence a large number of SDEs need to be included in the power system model, which will significantly influence the simulation speed. Also, the nature of the Gillespie method requires a large number of simulations on the same model to yield the mean trajectory as well as the envelope on variations. Therefore, adding any extra SDE to the existing set of SDEs will result in multiplying computing time by a factor of hundreds or even thousands.

In our previous works [12]-[14], a new semi-analytical approach for power system simulation and stability assessment has been proposed. That approach applies the Adomain decomposition method (ADM) to power system DAEs to derive a semi-analytical solution (SAS) for each state variable as an explicit function of symbolic variables including time, the initial system state and other selected parameters on the system condition; then each function is evaluated by plugging in values of its symbolic variables over consecutive small time windows to make up a desired simulation period so as to obtain the simulated trajectory of each state variable. Since the form of every SAS is a summation of finite terms for approximation, its evaluation can be fast and parallelized among terms. Thus, compared to traditional numerical integration based power system simulation, this semi-analytical approach decomposes the computation into offline derivation and online evaluation of an SAS and is better fit for online power system simulation and a parallel computing environment [14]. In fact, such a semi-analytical approach also suggests a viable, alternative paradigm for fast stochastic simulation. For example, early works by Adomian in the 1970s utilized the ADM to solve nonlinear SDEs [15] by embedding explicitly stochastic processes into the terms of an SAS.

For power system simulation and stability assessment in a stochastic manner, this paper proposes an approach as an extension of the ADM based approach proposed in [14]. Utilizing the semi-analytical nature of an SAS yielded by the ADM, this new approach embeds a stochastic model, e.g. a stochastic load model, into the SAS. Evaluation of an SAS with the stochastic model whose parameters are represented symbolically will not increase many computational burdens compared to evaluation of an SAS for deterministic simulation. Thus, an expected number of simulation runs for one single case are achieved by evaluating one SAS for the same number of times.

The rest of this paper is organized as follows: Section II presents the SDE model of a power system that integrates stochastic loads; Section III gives the ADM-based approach for solving the power system SDEs for stochastic simulation; Section IV uses a single-machine-infinite-bus (SMIB) system to compare the fundamental difference between the ADM-based approach and the Euler-Maruyama approach in mathematics; Section V validates the proposed approach using the IEEE 10-machine 39-bus system with the stochastic loads and outlines a potential application for power transmission distribution system cooperative control; finally, conclusions are drawn in Section VI.

## II. Power System SDE Model with Stochastic Loads

### A. Synchronous Generator Modeling

For a power system having $K$ synchronous generators, consider the 4th-order ordinary differential equation (ODE) model (1), the so-called "two-axis" model, to represent each generator with saliency ignored [16]. All generators are coupled through nonlinear algebraic equations (2) about the network.

$$\begin{cases} \dot{\delta}_k = \omega_k - \omega_R \\ \dot{\omega}_k = \dfrac{\omega_R}{2H_k}\left(P_{mk} - P_{ek} - D_k \dfrac{\omega_k - \omega_R}{\omega_R}\right) \\ \dot{e}'_{qk} = \dfrac{1}{T'_{d0k}}\left[E_{fdk} - e'_{qk} - \left(x_{dk} - x'_{dk}\right) i_{dk}\right] \\ \dot{e}'_{dk} = \dfrac{1}{T'_{q0k}}\left[-e'_{dk} + \left(x_{qk} - x'_{qk}\right) i_{qk}\right] \end{cases} \tag{1}$$

$$\begin{aligned} & E_k = e'_{dk}\sin\delta_k + e'_{qk}\cos\delta_k + j(e'_{qk}\sin\delta_k - e'_{dk}\cos\delta_k) \\ & I_{tk} = i_{Rk} + j i_{Ik} \overset{\text{def}}{=} \mathbf{Y}_k^{*}\mathbf{E} \\ & P_{ek} = e_{qk} i_{qk} + e_{dk} i_{dk} \\ & i_{qk} = i_{Ik}\sin\delta_k + i_{Rk}\cos\delta_k,\; i_{dk} = i_{Rk}\sin\delta_k - i_{Ik}\cos\delta_k \\ & e_{qk} = e'_{qk} - x'_{dk} i_{dk},\; e_{dk} = e'_{dk} + x'_{qk} i_{qk} \end{aligned} \tag{2}$$

In (1) and (2), $\omega_R$ is the rated angular frequency; $\delta_k$ , $\omega_k$, $H_k$ and $D_k$ are respectively the rotor angle, rotor speed, inertia and damping coefficient of the machine $k$; $\mathbf{Y}_k$ is the $k$th row of the reduced admittance matrix $\mathbf{Y}$; $\mathbf{E}$ is the column vector of all generator's electromotive forces (EMFs) and $E_k$ is the $k$th element; $P_{mk}$ and $P_{ek}$ are the mechanical and electric powers; $E_{fdk}$ is the internal field voltage; $e'_{qk}$, $e'_{dk}$, $i_{qk}$, $i_{dk}$, $T'_{q0k}$, $T'_{d0k}$, $x_{qk}$, $x_{dk}$, $x'_{qk}$ and $x'_{dk}$ are transient voltages, stator currents, open-circuit time constants, synchronous reactances and transient reactances in $q$- and $d$-axes, respectively.

### B. Stochastic Load Modeling

A stochastic model can be built based on analysis on real data and assumptions on probabilistic characteristics of the stochastic variables. Traditionally, uncertainties in loads of a power system are ignored in time-domain simulation for the sake of simplicity. However their stochastic behaviors are well-recognized in [17]. Taking stochastic loads into consideration will enable more realistic power system stability assessment.

This paper uses the Ornstein-Uhlenbeck process in [18] to model the stochastic variations of a load in these SDEs:

$$\dot{\mathbf{y}}_{\mathrm{PL}} = -\mathbf{a}_{\mathrm{P}} \circ \mathbf{y}_{\mathrm{PL}} + \mathbf{b}_{\mathrm{P}} \circ \mathbf{W}(t) \tag{3}$$

$$\dot{\mathbf{y}}_{\mathrm{QL}} = -\mathbf{a}_{Q} \circ \mathbf{y}_{\mathrm{QL}} + \mathbf{b}_{Q} \circ \mathbf{W}(t) \tag{4}$$

where $\mathbf{W}$(t) is the white noise vector whose dimension equals the number of load buses, $\mathbf{a}$ and $\mathbf{b}$ parameters are drifting and diffusion parameters of the SDEs, operator " $\circ$ " is the Hadamard Product, i.e., element-wise multiplication, and $\mathbf{y}_{\mathrm{PL}}$ and $\mathbf{y}_{\mathrm{QL}}$ are the stochastic variations in normal distributions.

The stochastic dynamic of the load is therefore modeled by

$$\mathbf{P}_{\mathrm{L}} = \mathbf{P}_{\mathrm{L0}} + \mathbf{y}_{\mathrm{PL}} \tag{5}$$

$$\mathbf{Q}_{\mathrm{L}} = \mathbf{Q}_{\mathrm{L0}} + \mathbf{y}_{\mathrm{QL}} \tag{6}$$

where $\mathbf{P}_{\mathrm{L0}}$ and $\mathbf{Q}_{\mathrm{L0}}$ are the mean values of the active and reactive loads, respectively.

Periodicities and autocorrelations have been observed in historical data of loads on the daily basis. However, in the time frame of seconds, loads at different substations have much lower autocorrelations. Refer to [7], this paper sets the drifting parameter on the autocorrelations of loads as 0.5 p.u./s.

## III. Proposed ADM-based Approach for Solving Power System SDEs

### A. Modeling Stochastic Variables

Consider $S$ stochastic variables $y_1(t),\cdots,y_S(t)$, which could be stochastic loads following $S$ different distributions. Each $y_i(t)$ can be transformed by function $g_i(\cdot)$ in (7) from some $\varepsilon_i$ in a normal distribution. For example, if $y_i(t)$ is a load represented by a normal distribution with certain mean value, then $\varepsilon_i$ specifies a zero-mean normal distribution as in (9) and $g_i(\cdot)$ shifts it to around the desired mean value like in (5) and (6).

$$\mathbf{y}(t) = \begin{bmatrix} g_1(\varepsilon_1) & g_2(\varepsilon_2) & \cdots & g_S(\varepsilon_S) \end{bmatrix}^T \tag{7}$$

The Ornstein-Uhlenbeck process is utilized to generate each $\varepsilon_i$ from (8).

$$\dot{\boldsymbol{\varepsilon}}(t) = -\mathbf{a} \circ \boldsymbol{\varepsilon}(t) + \mathbf{b} \circ \mathbf{W}(t) \tag{8}$$

where
$$\boldsymbol{\varepsilon}(t) = \left[\varepsilon_1(t) \quad \varepsilon_2(t) \quad \cdots \quad \varepsilon_S(t)\right]^T$$
$$\mathbf{a}(t) = \left[a_1(t) \quad a_2(t) \quad \cdots \quad a_S(t)\right]^T$$
$$\mathbf{b}(t) = \left[b_1(t) \quad b_2(t) \quad \cdots \quad b_S(t)\right]^T$$
$$\varepsilon_i \sim \mathcal{N}(0, b_i^2/2a_i) \quad i = 1, 2, \cdots, S \tag{9}$$

### B. Solving SDEs Using the ADM

Consider a nonlinear system modeled by SDE (10) having $M$ deterministic state variables $x_1$, …, $x_M$, such as the state variables of generators, exciters and speed governors, and $S$ stochastic variables $y_1$, …, $y_S$.

$$\dot{\mathbf{x}}(t) = \mathbf{f}(\mathbf{x}(t), \mathbf{y}(t)) \tag{10}$$
$$\mathbf{x}(t) = \left[x_1(t) \quad x_2(t) \quad \cdots \quad x_M(t)\right]^T$$
$$\mathbf{f}(\cdot) = \left[f_1(\cdot) \quad f_2(\cdot) \quad \cdots \quad f_M(\cdot)\right]^T$$

To solve $\mathbf{x}(t)$, the procedure in [14] can be used. First, apply Laplace transformation to (10) to obtain

$$\mathcal{L}[\mathbf{x}] = \frac{\mathbf{x}(0)}{s} + \frac{\mathcal{L}[\mathbf{f}(\mathbf{x}, \mathbf{y})]}{s} \tag{11}$$

Then use (13) and (14) to calculate the Adomian polynomials under the assumption of (12),

$$\mathbf{x}(t) = \sum_{n=0}^{\infty} \mathbf{x}_n(t) \tag{12}$$
$$f_k(\mathbf{x}, \mathbf{y}) = \sum_{n=0}^{\infty} A_{k,n}(\mathbf{x}_0, \mathbf{x}_1, \ldots \mathbf{x}_n, \mathbf{y}) \tag{13}$$
$$A_{k,n} = \frac{1}{n!}\left[\frac{\partial^n}{\partial \lambda^n} f_k\left(\sum_{k=0}^{n} \mathbf{x}_k \lambda^k, \mathbf{y}\right)\right]\Bigg|_{\lambda=0} \tag{14}$$

Recursive formulas (15) and (16) can be derived by matching terms of $\mathbf{x}(t)$ and $\mathbf{f}(\cdot)$:

$$\mathcal{L}\left[\mathbf{x}_0\right] = \mathbf{x}(0)/s \tag{15}$$
$$\mathcal{L}\left[\mathbf{x}_{n+1}\right] = \mathcal{L}\left[\mathbf{A}_n\right]/s \quad n \geq 0 \tag{16}$$
$$\text{where } \mathbf{A}_n = \left[A_{1,n}, A_{2,n}, \cdots, A_{M,n}\right]^T$$

The next step is to apply inverse Laplace transform to both sides of (15) and (16) to calculate the $N$-th order SAS of (10):

$$\mathbf{x}^{SAS}(t, \mathbf{y}) = \sum_{n=0}^{N} \mathbf{x}_n(t, \mathbf{y}) \tag{17}$$

In the resulting SAS, stochastic variables in $\mathbf{y}$ appear explicitly as symbolic variables.

## IV. Comparison between the Euler-Maruyama Approach and ADM Based Approach

This section applies both the Euler-Maruyama approach and the proposed ADM-based approach to the SMIB system with a stochastic load shown in Fig. 1 to illustrate the fundamental difference between the two approaches.

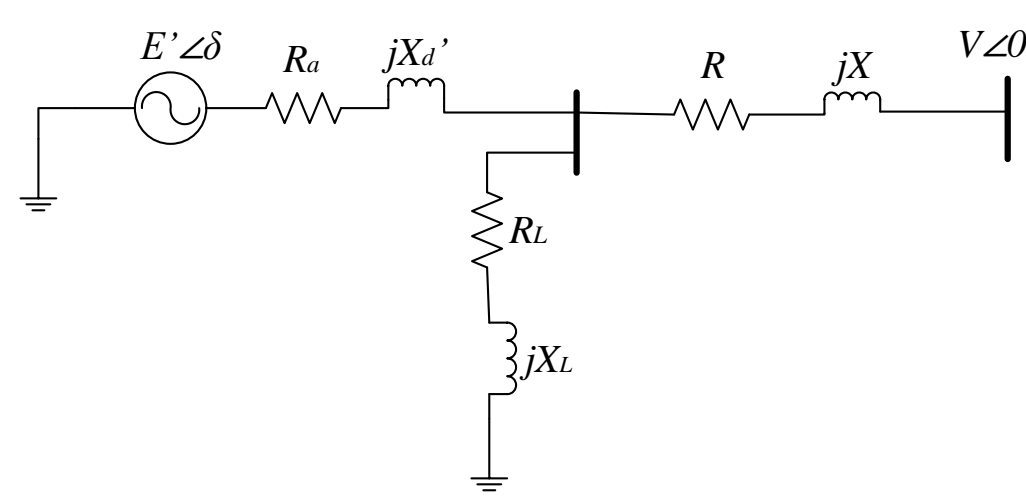

Fig. 1. SMIB system with constant impedance load at generator bus.

The stochastic load is connected to the generator bus and has its resistance $R_L$ and reactance $X_L$ modeled by stochastic variables. Thus, the whole system is now modeled by ODEs (18a), (18b) and SDEs (18c), (18d).

$$\begin{cases} \dot{\delta} = \omega - \omega_R & \text{(18a)} \\ \dot{\omega} = \dfrac{\omega_R}{2H}\left(-D\dfrac{\omega - \omega_R}{\omega} + P_m - \left(k_3 + \dfrac{E'V}{k_1 k_2}\left(k_4\cos(\delta) + k_5\sin(\delta)\right)\right)\right) & \text{(18b)} \\ \dot{R}_L = -a_1 R_L + b_1 W(t) & \text{(18c)} \\ \dot{X}_L = -a_2 X_L + b_2 W(t) & \text{(18d)} \end{cases}$$

where,

$$G_L + jB_L = \frac{1}{R_L + jX_L} \tag{19a}$$
$$G_S + jB_S = \frac{1}{R_a + jX_d'} \tag{19b}$$
$$G_R + jB_R = \frac{1}{R + jX} \tag{19c}$$
$$k_1 = \frac{(G_L + G_R + G_S)^2}{(B_L + B_R + B_S)} + (B_L + B_R + B_S) \tag{19d}$$
$$k_2 = (G_L + G_R + G_S) + \frac{(B_L + B_R + B_S)^2}{(G_L + G_R + G_S)} \tag{19e}$$
$$k_3 = E'^2\left(\frac{G_S(B_L + B_R) + B_S(G_L + G_R)}{k_1} - \frac{B_S(B_L + B_R) - G_S(G_L + G_R)}{k_2}\right) \tag{19f}$$
$$k_4 = -k_2(B_S G_R + G_S B_R) + k_1(B_S B_R - G_S G_R) \tag{19g}$$
$$k_5 = -k_2(B_S B_R - G_S G_R) - k_1(B_S G_R + G_S B_R) \tag{19h}$$

In (19), $G_S$, $B_S$, $G_R$, $B_R$, $G_L$, $B_L$ are the conductances and susceptances at the generator sending side, the infinite bus receiving side and the load side, respectively. Since $R_L$ and $X_L$ change stochastically, $G_L$ and $B_L$ can not be treated as constants.

The variances of $R_L$ and $X_L$ depend on the values of drifting parameters $a_1$ and $a_2$ and diffusion parameters $b_1$ and $b_2$, respectively.

To find the SAS of this system, the first step is to apply ADM to the ODEs (18a) and (18b). The resulting 2nd order SAS for rotor speed $\omega$ is,

$$\omega(t) = \sum_{n=0}^{2} \omega_n(t) \tag{20}$$

where,

$$\omega_0(t) = \omega(0) \tag{21}$$

$$\omega_1(t) = -\frac{t\omega_R}{2H}\left[\frac{D(\omega(0)-\omega_R)}{\omega_R} - P_m + k_3 + \frac{k_4E'V}{k_1k_2}\cos(\delta(0)) + \frac{k_5E'V}{k_1k_2}\sin(\delta(0))\right] \tag{22}$$

$$\begin{aligned}\omega_2(t) = \frac{t^2\omega_R}{8H^2}\Bigg[&\frac{D^2(\omega(0)-\omega_R)}{\omega_R} \\ &+D(-P_m + k_3 + \frac{k_4E'V}{k_1k_2}\cos(\delta(0)) + \frac{k_5E'V}{k_1k_2}\sin(\delta(0))) \\ &+2H\omega_R\frac{k_5E'V}{k_1k_2}\cos(\delta(0)) - 2H\omega_R\frac{k_4E'V}{k_1k_2}\sin(\delta(0)) \\ &-2H(\omega(0)-\omega_R)\frac{E'V}{k_1k_2}\cos(\delta(0))\Bigg]\end{aligned} \tag{23}$$

Once the SAS of the system's ODEs is derived, the SAS of the SDEs can be derived and incorporated into it.

For example, the 2nd order SAS of $R_L$ can be derived using ADM as,

$$R_L(t) = \sum_{n=0}^{2} R_{L,n}(t) \tag{24}$$

where

$$R_{L,0}(t) = R_L(0) + b_1B(t) \tag{25}$$

$$R_{L,1}(t) = -a_1R_L(0)t - a_1b_1\int_0^t B(s_1)ds_1 \tag{26}$$

$$R_{L,2}(t) = a_1^2R_L(0)\frac{t^2}{2!} + a_1^2b_1\int_0^t\int_0^{s_1} B(s_2)ds_2ds_1 \tag{27}$$

$B(t)$ is the Brownian motion starting at origin and $dB(t)=W(t)dt$.

The 2nd order SAS of $X_L$ can be derived in similar manner as in (24) to (27).

To derive the SAS of the entire system considering both the ODEs and SDEs, replace the symbolic variables in the DEs' SAS representing the stochastic variables with the SDEs' SAS, i.e., the 2nd order SAS of the system (18) can be derived by replacing the symbolic variables $R_L$ and $X_L$ in (20) with their SASs.

For some forms of SDE, the analytical solution may exist. In such cases, the SDEs' analytical solution instead of the SAS also can be incorporated into the ODEs' SAS to derive the SAS of the entire system.

For example, the general expression of the SAS terms of (18c) can be written as,

$$R_{L,n}(t) = (-1)^n a_1^n R_L(0)\frac{t^n}{n!} + (-1)^n a_1^n b_1 \int_0^t\int_0^{s_1}\cdots\int_0^{s_{n-1}} B(s_n)ds_n\ldots ds_2ds_1 \tag{28}$$

Therefore the infinite order SAS of (20c) is,

$$R_L(t) = R_L(0)\sum_{i=0}^{\infty}\frac{(-a_1t)^i}{i!} + b_1B(t) + b_1\sum_{i=1}^{\infty}(-a_1)^i\int_0^t\int_0^{s_1}\cdots\int_0^{s_{i-1}} B(s_i)ds_i\ldots ds_2ds_1 \tag{29}$$

Apply Maclaurin expansion of an exponential function and lemma 2.3 in [19] to (29), the solution becomes,

$$R_L(t) = R_L(0)e^{-a_1t} + b_1B(t) - a_1b_1\int_0^t e^{a_1s-a_1t}B(s)ds \tag{30}$$

Then apply the integration by parts formula,

$$\int_0^t e^{a_1s}dB(s) = e^{a_1t}B(t) - \int_0^t a_1e^{a_1s}B(s)ds \tag{31}$$

The close form solution can be found as,

$$R_L(t) = e^{-a_1t}[R_L(0) + b_1\int_0^t e^{a_1s}dB(s)] \tag{32}$$

In this case the symbolic variable $R_L$ in (20) can be replaced by (32) instead of (24).

On the other hand, for the Euler-Maruyama approach [20][21], since the deterministic model described by (18a) and (18b) does not permit a close form solution, the sample trajectories of (18) have to be numerically computed. The numerical scheme for $R_L$ is shown in (33) and the same scheme also applies to $X_L$.

$$R_{L,n+1}^{(\Delta t)} = R_{L,n}^{(\Delta t)} + a_1R_{L,n}^{(\Delta t)}\Delta t + b_1R_{L,n}^{(\Delta t)}\Delta W \tag{33}$$

In practice the value of $\Delta W$ is dependent of the step size $\Delta t$ for integration.

$$\Delta W \sim \mathcal{N}(0, \Delta t^{1/2}) \tag{34}$$

## V. Case Studies

The proposed ADM-based approach is tested on the IEEE 10-machine 39-bus New England system as shown in Fig. 2. All loads are assumed to change stochastically while all generators are represented by deterministic models. In each case study, the stochastic simulation result by the Euler-Maruyama approach is used as the benchmark, and the 2nd order SASs (i.e. $N$=2) are used and evaluated every 0.001 s. The value of each stochastic variable is changed every 0.1 s. To make the studies repeatable and the comparisons fair, a same pseudo random number generating scheme is used for both approaches. For each case, 100 sample trajectories are generated. The fault applied in cases A and B is a self-clearing 4-cycle 3-phase fault at bus 1. All simulations are performed in MATLAB R2017b on a laptop computer with an Intel Core i5-6300U 2.40GHz CPU and 4 GB RAM.

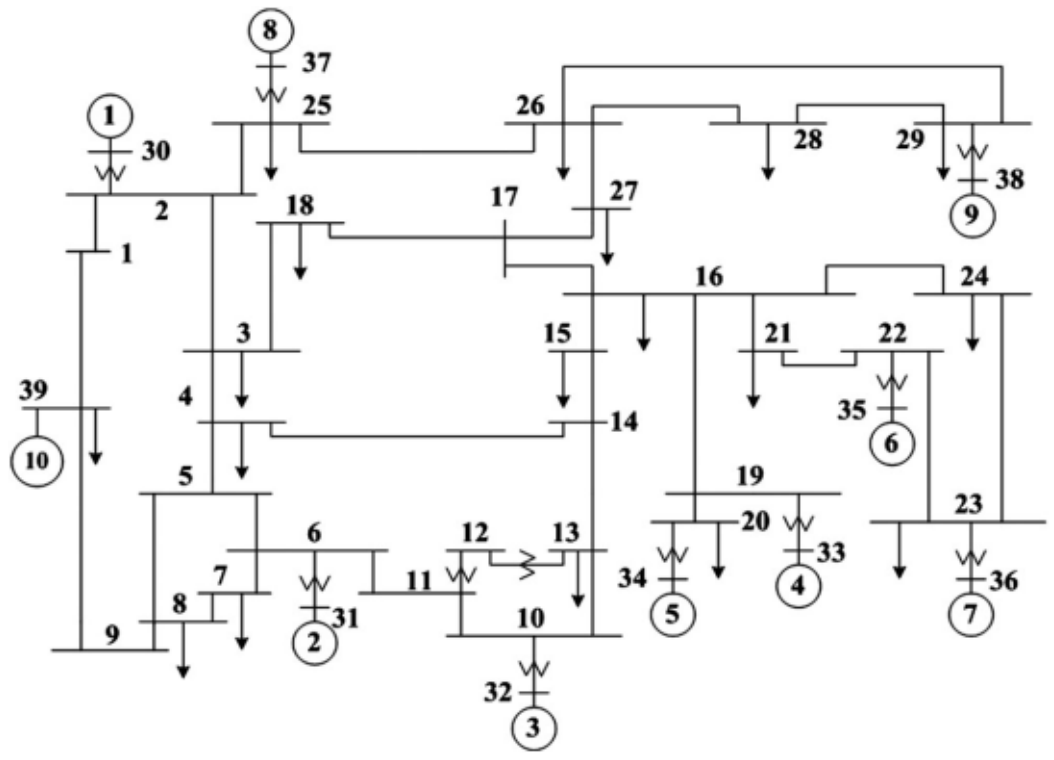

Fig. 2. IEEE 10-machine 39-bus system.

### A. Stochastic Loads with Low Variances

In the first case, the variances of the loads are 0.1% of their mean values. The results of 100 runs from the ADM-based approach are shown in Fig. 3. Among all the generators, generator 1 has the shortest the electrical distance to bus 1, hence its rotor speed deviation from 377 rad/s is presented in the following results.

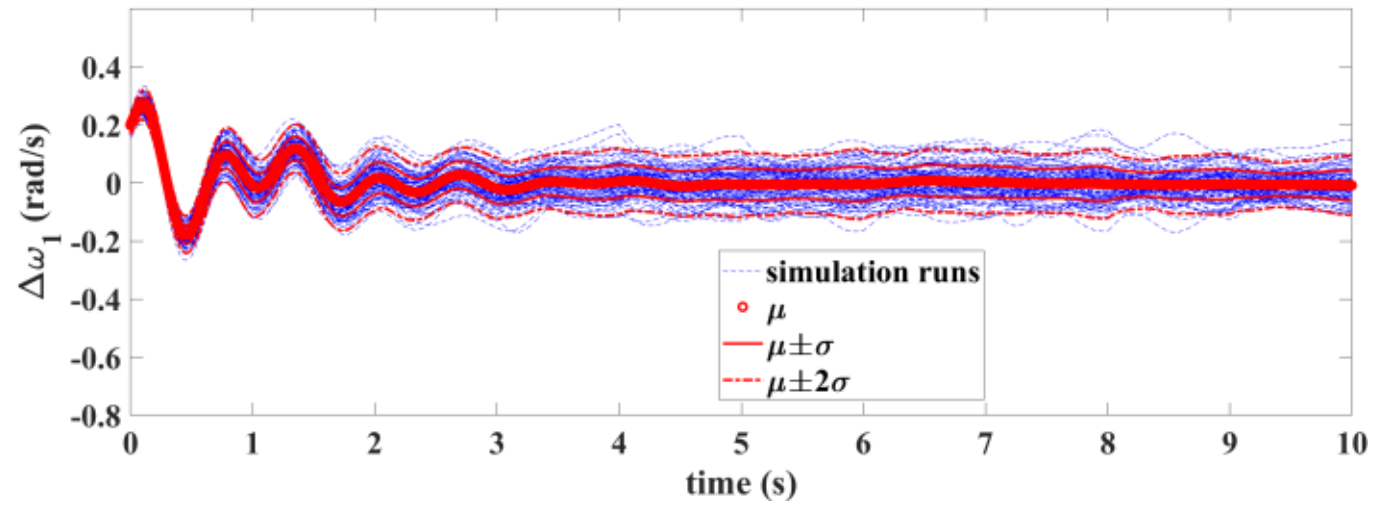

Fig. 3. ADM-based stochastic simulation results of generator 1 rotor speed deviation (low load variances case).

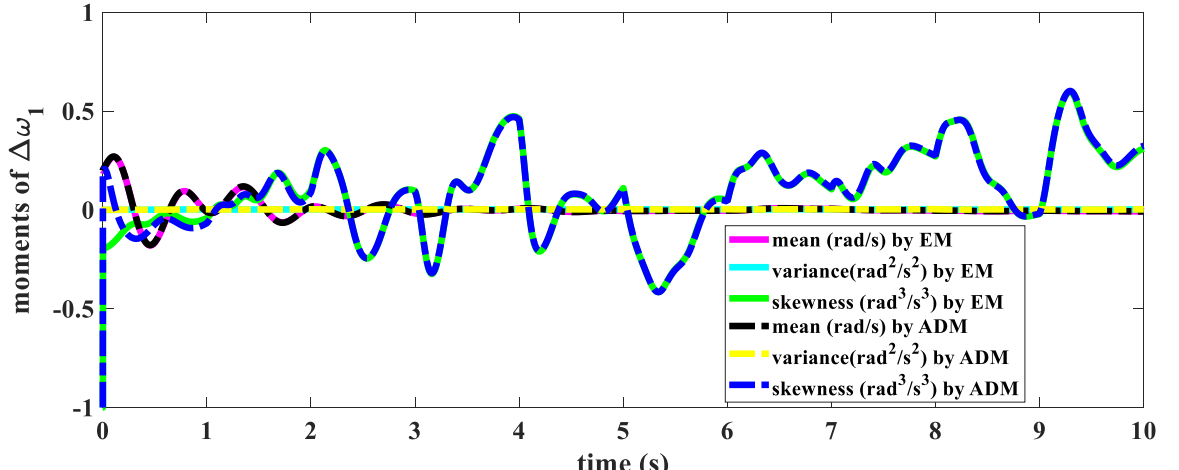

Fig. 4. Mean, variance and skewness comparisons of the generator 1 rotor speed deviation from the ADM-based approach and the Euler-Maruyama approach (low load variance case).

As shown in Fig. 3, under the assumption of low load variances, the mean trajectory of 100 trajectories is a sufficient representation of the system dynamic. The $\mu$+$\sigma$ and $\mu$+2$\sigma$ envelopes (which encloses 68% and 95% of the sample trajectories, respectively) are very close to the mean trajectory. To benchmark the accuracy of the ADM-based approach, the mean, variance and skewness of the trajectories from the proposed approach are compared with those from Euler-Maruyama approach. The comparisons show good agreement in Fig. 4.

## *B. Stochastic Loads with High Variances*

In the second case, the variances of the loads are increased to 1% of their mean values. As the uncertainty in the system increases, the mean trajectory can no longer represent the system dynamic.

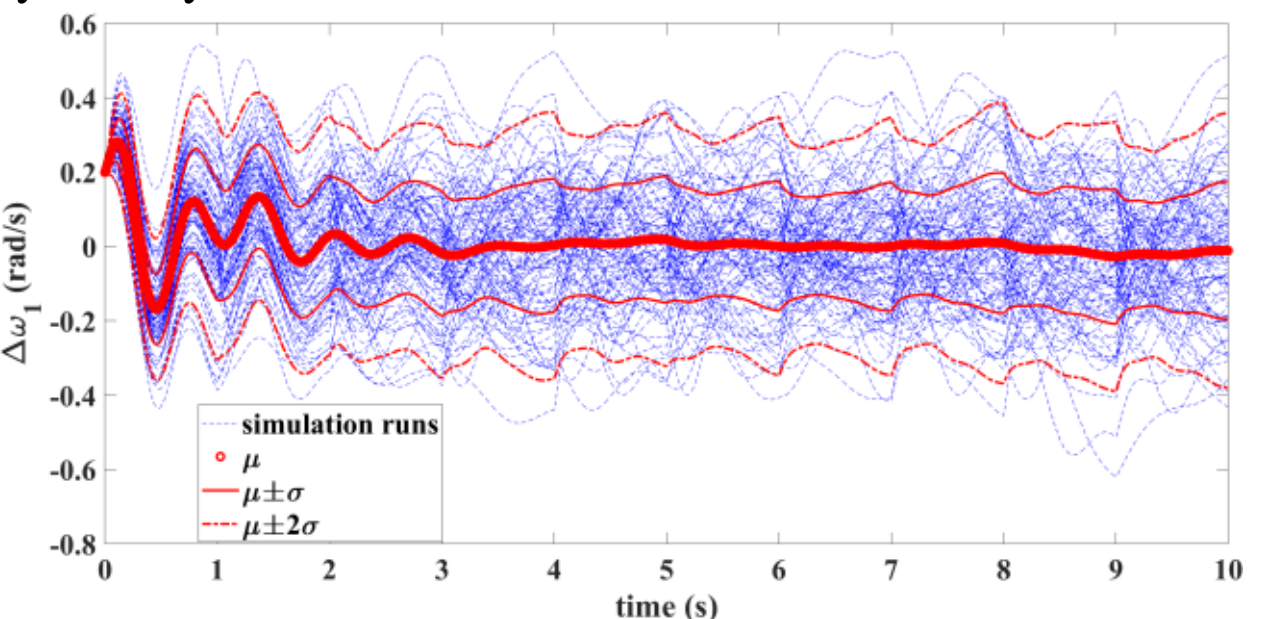

Fig. 5. ADM-based stochastic simulation results of generator 1 rotor speed deviation (high load variances case).

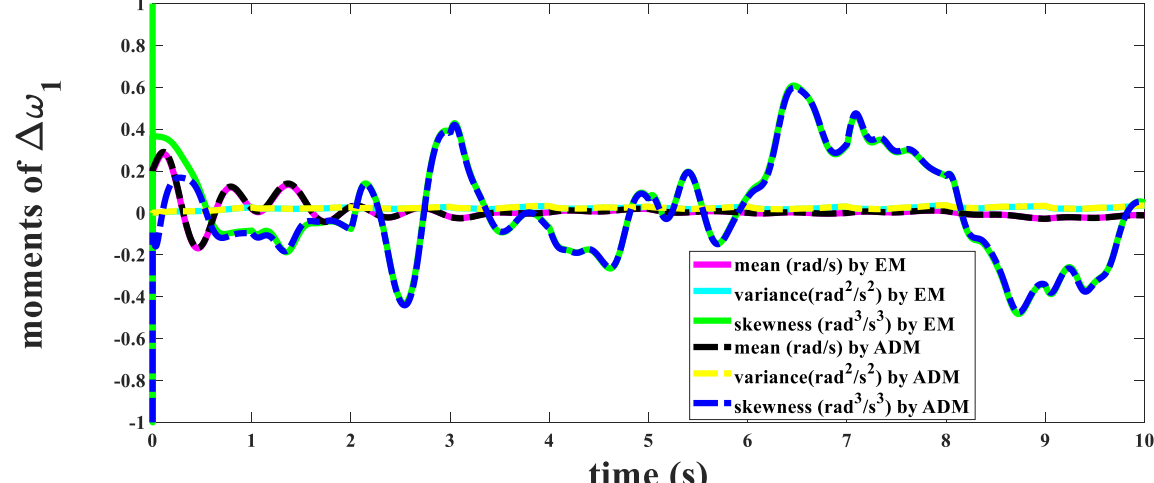

Fig. 6. Mean, variance and skewness comparisons of the generator 1 rotor speed deviation from the ADM-based approach and the Euler-Maruyama approach (high load variance case).

As shown in Fig. 5, the $\mu$+$\sigma$ and $\mu$+2$\sigma$ envelopes are noticeably apart from the mean trajectory, they even show an undamped behavior. It implies that the probability of the system becomes marginal unstable is no long negligible. That justifies the necessity of using stochastic load models to study the stability of power systems under random perturbations due to stochastic loads highly penetrating into the grid.

The accuracy of the proposed approach is not affected by the increase of uncertainties. As shown in Fig. 6, the mean, variance and skewness from proposed approach still match those from the Euler-Maruyama approach very well.

## *C. Comparison on Time Performances*

The time performances for the ADM-based approach and the Euler-Maruyama approach are compared in TABLE I, from which the ADM-based approach only takes 75% of the time cost of the Euler-Maruyama approach. The advantage of the ADM-based approach in time performance is more prominent when many simulation runs are required. As discussed in [14], the ADM-based approach is inherently suitable for parallel implementation, which could help further improve the time performance if high-performance parallel computers are available.

TABLE I
TIME PERFORMANCE COMPARISON OF STOCHASTIC LOAD CASES

| Methods | Time costs (s) |
|---|---|
| Euler-Maruyama 100 runs | 424.8 |
| ADM 100 runs | 317.0 |

## *D. EMS-DMS Cooperative Control Informed by the ADM-based Stochastic Simulation*

Energy management system (EMS) and distribution management system (DMS) are two essential components that inform the operation of power transmission and distribution systems. They are traditionally considered independent, but more and more efforts are being made to integrate them for more reliable and efficient control and planning. The faster stochastic simulation enabled by ADM provides an opportunity for EMS and DMS to have transactional communication and achieve cooperative control.

Under deterministic assumption, to mitigate instability, the most common remedial action is load shedding, but the economic impact of load shedding cannot be overlooked. With stochastic simulation, the uncertainty of load becomes a new attribute based on which new control strategies can be designed.

The following case demonstrates a scenario where the DMS has sufficient capability to adjust load scheduling, renewable energy generation and energy storage operations to reduce the uncertainty of the aggregative behavior of the loads under a distribution feeder [22]. Such an uncertainty reduction strategy is associated with certain cost therefore undesirable under normal operation when the power transmission system can maintain its stability.

However, when potential instability is indicated by the stochastic simulation performed by EMS, instead of apply load shedding, load uncertainty reduction strategy can be applied to alleviate the stress on the system, which comes with a much lower cost than load shedding. As shown in Fig. 7, a 30-cycle self-clearing 3-phase fault is added to bus 1. With all loads having 2% of mean value variance, the system cannot be treated as an autonomous system any more. Its stays within some domain of attraction of the stable equilibrium for first several swings, but then its oscillation grows to approach the boundary of the stability domain. In Fig. 8, the DMSs at all the load buses receive the control signal from the EMS to apply load uncertainty reduction strategy at t = 6 s. The system is stabilized after that because of the cooperative control of EMS and DMSs. The foundation of such uncertainty based control action is the advance SDE simulation approaches such as the ADM-based approach.

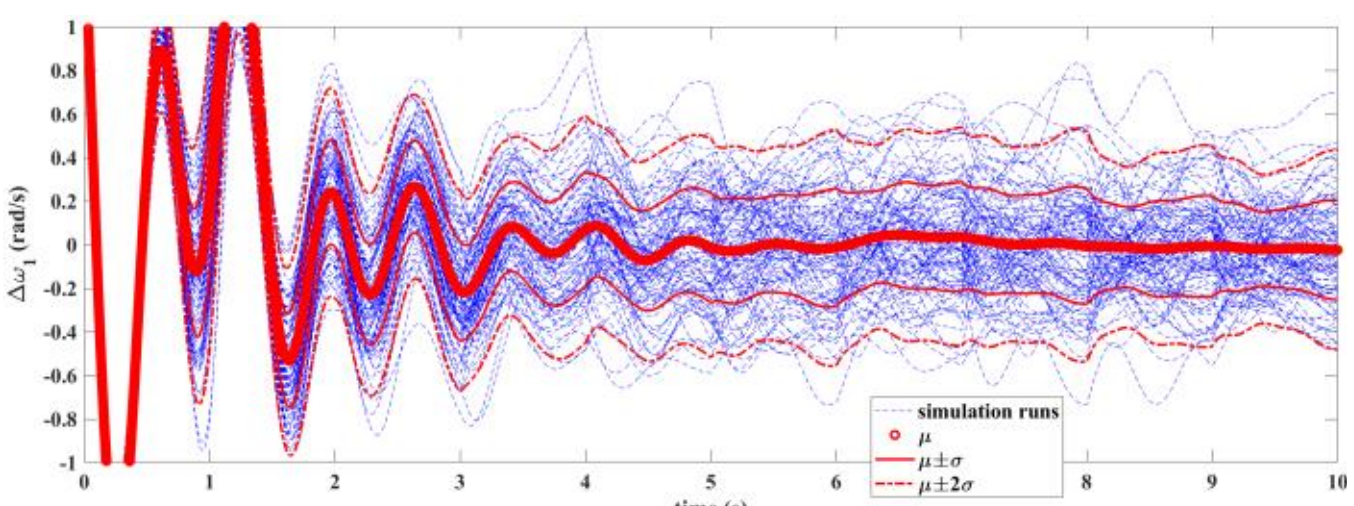

Fig. 7. Results of the ADM-based stochastic simulation without control.

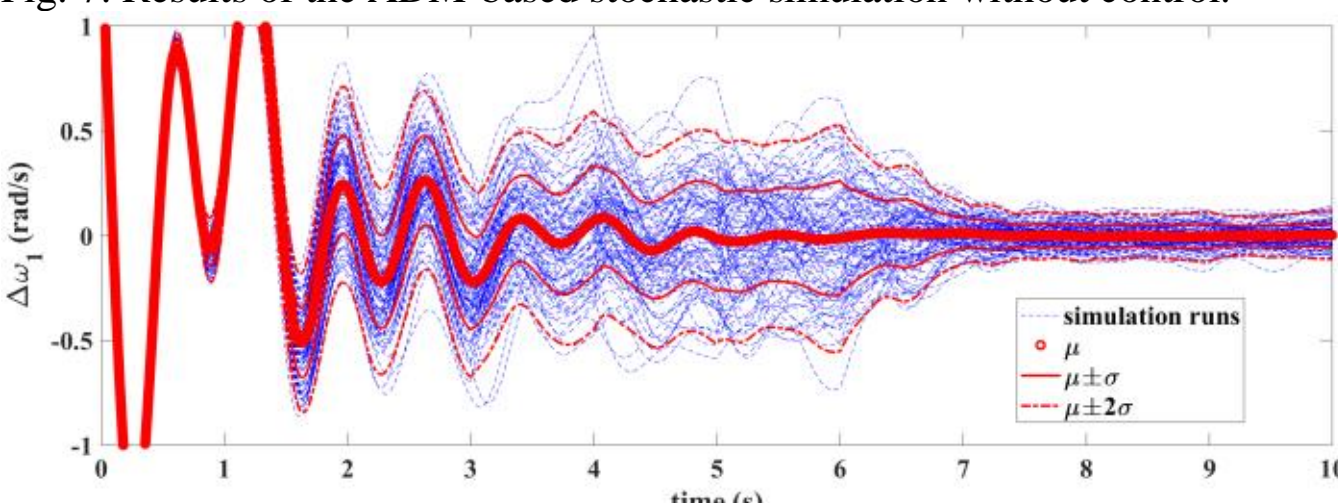

Fig. 8. Results of the ADM-based stochastic simulation with the EMS-DMS cooperative control taking effect at t = 6 s.

## VI. Conclusion

This paper proposes an alternative approach for stochastic simulation and stability assessment of power systems. Using the SAS derived from the ADM, the stochastic effects from load uncertainties can be taken into considerations. The results from the proposed approach are benchmarked with those from the Euler-Maruyama approach. Since the evaluation of SASs is faster than the integration with the Euler-Maruyama approach, the proposed approach has an obviously advantage in time performance. This is critical when a large number of simulation runs need to be performed for simulating stochastic behaviors of a future power grid having a high penetration of DERs. The simulation results on different levels of stochastic loads show that when the level of load uncertainty is low, the deterministic simulation is still trustworthy compared to the mean-value trajectory from stochastic simulation, but, once the level of load uncertainty becomes high, the mean-value trajectory no longer represents the true behavior of the system. This paper also demonstrates that modeling uncertainty in power system provides the opportunity to study uncertainty based control strategies.